# Resonant collisional shielding of reactive molecules using electric fields

Kyle Matsuda[1]*, Luigi De Marco[1], Jun-Ru Li[1], William G. Tobias[1], Giacomo Valtolina[1], Goulven Quéméner[2], and Jun Ye[1]*

[1] JILA, National Institute of Standards and Technology and Department of Physics, University of Colorado, Boulder, Colorado 80309, USA.

[2] Université Paris-Saclay, CNRS, Laboratoire Aimé Cotton, 91405, Orsay, France.

**Abstract:** Full control of molecular interactions, including reactive losses, would open new frontiers in quantum science. Here, we demonstrate extreme tunability of chemical reaction rates by using an external electric field to shift excited collision channels of ultracold molecules into degeneracy with the initial collision channel. In this situation, resonant dipolar interactions mix the channels at long range, dramatically altering the intermolecular potential. We prepare fermionic potassium-rubidium (KRb) molecules in their first excited rotational state and observe a three orders-of-magnitude modulation of the chemical reaction rate as we tune the electric field strength by a few percent across resonance. In a quasi-two-dimensional geometry, we accurately determine the contributions from the three lowest angular momentum projections of the collisions. Using the resonant features, we shield the molecules from loss and suppress the reaction rate by up to an order of magnitude below the background value, realizing a long-lived sample of polar molecules in large electric fields.

**Main text:**

Controlling chemical reactions and collisions has been a central focus of work on cold and ultracold molecules (*1–6*). Progress in cooling and trapping molecules has led to exciting advances in this area, including the precise characterization of scattering resonances (*7–9*), the observation of atom-molecule (*10*) and molecule-molecule (*11–13*) cold collisions, and the synthesis of new chemical species (*14*). In particular, ultracold polar molecules, for which both internal and external degrees of freedom are controlled, present unique opportunities (*15–26*). At ultralow temperatures, small perturbations to the long-range intermolecular potential, while negligible compared to chemical bonding energy scales, can vastly exceed the kinetic energy of the colliding molecules and thus strongly alter the rate of chemical reactions at close range (*3*). This sensitivity, combined with the rich structure of polar molecules and their tunability using external electromagnetic fields, suggests the exciting possibility of precisely controlling reactions (*1, 4–6*). In addition to providing insights about fundamental chemical processes (*27, 28*), such control would aid in the production of quantum-degenerate molecular gases (*29–31*) and facilitate precision measurements (*32*) or studies of many-body physics (*33*) in these systems.

Applying an electric field $\vec{E}$ strongly modifies the reaction rates of ultracold polar molecules via anisotropic dipolar interactions. In three dimensions (3D), attractive head-to-tail collisions lead to rapid losses (*34, 35*), which scale for fermions as $d^6$ in the induced dipole moment $d$ (*36*). If fermionic molecules are instead trapped in a quasi-two-dimensional (quasi-2D) geometry with $\vec{E}$ along the strongly confined direction, only repulsive side-to-side collisions are allowed, suppressing losses while enhancing the elastic collision rate (*37, 38*). For reactive ground-state KRb molecules, we recently achieved a ratio of elastic to inelastic collisions exceeding 100 using this approach (*31*).

Here, we demonstrate a striking effect of the electric field on molecular collisions—chemical reaction rates in an ultracold gas of molecules are sharply varied by three orders of magnitude near particular values of the field strength $|\vec{E}|$. These values occur where higher rotationally excited channels become degenerate with the initial collision channel, inducing resonant dipolar interactions that profoundly alter the long-range potential and hence the reaction rate (*39*). This mechanism was first proposed by Avdeenkov *et al.* (*40*), and related theory was subsequently extended to a wide variety of bosonic and fermionic species of experimental interest (*39, 41, 42*). In this work, we experimentally demonstrate this effect for the first time.

Specifically, we prepare ultracold fermionic $^{40}$K$^{87}$Rb molecules in the $|N, m_N\rangle = |1,0\rangle$ state, with $N$ the rotational angular momentum and $m_N$ its projection onto the axis of $\vec{E}$. We observe a drastic change in the two-body reactive loss rate near two field strengths, $|\vec{E}_1| = 11.72$ kV/cm and $|\vec{E}_2| = 12.51$ kV/cm, where the energies of the $|0,0\rangle|2,\pm1\rangle$ and $|0,0\rangle|2,0\rangle$ collision channels (respectively) cross the energy of $|1,0\rangle|1,0\rangle$ (Fig. 1A). Near these degeneracies (Fig. 1B), dipolar interactions strongly couple the two crossing channels at intermolecular separations $r \sim r_0$, where $r_0$ is the radius of the $p$-wave centrifugal barrier ($r_0 = 270 a_0$ for ground-state KRb, with $a_0$ the Bohr radius) (*43*). This coupling leads to a large effective van der Waals interaction $V_{vdw}$, which can be attractive or repulsive depending on the sign of the detuning of $|\vec{E}|$ from resonance. Figure 1C shows the calculated adiabatic energy curves for two colliding KRb molecules at two fields near $|\vec{E}_2|$. For $|\vec{E}| > |\vec{E}_2|$ (orange line), the energy of the $|1,0\rangle|1,0\rangle$ channel is higher than that of the $|0,0\rangle|2,0\rangle$ channel. This creates a repulsive $V_{vdw}$ interaction ($\sim 300$ µK) that is three orders of magnitude larger than the typical collision energy set by the temperature of the gas (250 nK). In this case, molecules are shielded from reactive losses, since they cannot meet at short range except by tunneling through the barrier (which occurs with a very low probability). Conversely, for $|\vec{E}| < |\vec{E}_2|$ (green line), the energy of $|1,0\rangle|1,0\rangle$ is lower than that of $|0,0\rangle|2,0\rangle$ and $V_{vdw}$ is attractive, resulting in an enhanced loss rate.

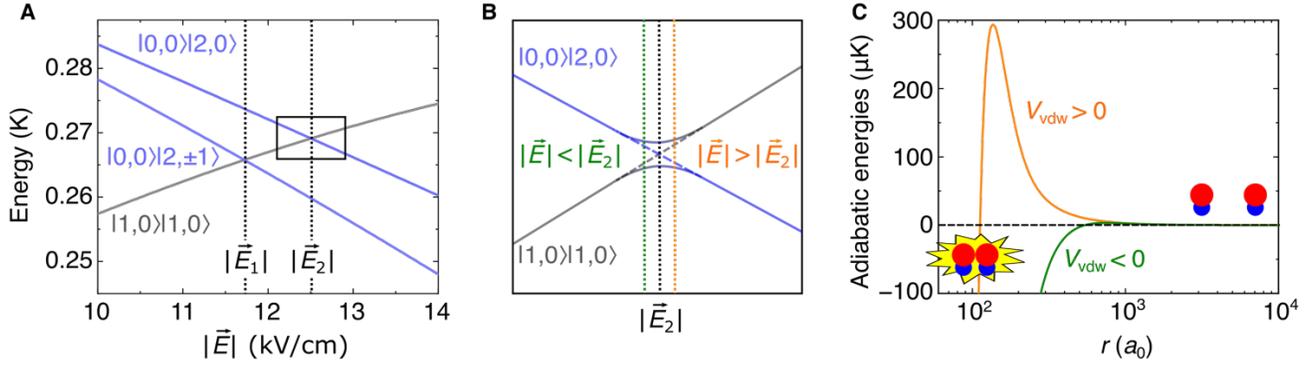

**Fig. 1. Electric field induced shielding.** (**A**) Energies of the relevant collision channels as a function of $|\vec{E}|$. Reactive losses of molecules colliding in the $|1,0\rangle|1,0\rangle$ channel can be sharply tuned near two values of the applied electric field, $|\vec{E}_1| = 11.72$ kV/cm and $|\vec{E}_2| = 12.51$ kV/cm, where the energy crosses the $|0,0\rangle|2,\pm1\rangle$ and $|0,0\rangle|2,0\rangle$ thresholds, respectively. (**B**) Qualitative picture of the region near $|\vec{E}_2|$. For large molecular separations $r \gg r_0$ (dashed lines), the two channels are not coupled. For separations $r \sim r_0$ (solid lines), an avoided crossing is opened due to dipolar interactions between the scattering channels. As two $|1,0\rangle$ molecules collide at a fixed $|\vec{E}|$ greater than (less than) $|\vec{E}_2|$, the avoided crossing results in an effective van der Waals repulsion (attraction). The same argument applies near $|\vec{E}_1|$. (**C**) Adiabatic energy curves for colliding KRb molecules in $|1,0\rangle$ at $|\vec{E}| = 12.504$ kV/cm (green) and 12.670 kV/cm (orange). The dashed line shows the average collision energy of 250 nK. For $|\vec{E}| > |\vec{E}_2|$, the barrier $V_{vdw}$ can be much higher than the collision energy, and the molecules must tunnel through the barrier in order to react at short range.

This effect is akin to a Förster resonance, for example between Rydberg atoms (*44–46*), in which $|\vec{E}|$ is tuned to create degeneracies between pairs of dipole-coupled states, resulting in resonant energy transfer. A key difference is the much smaller dipole moment of molecules compared to Rydberg atoms. Consequently, colliding molecules experience an adiabatic increase in the dipolar interaction energy as they approach. We stress that this is not a conventional scattering resonance arising from the presence of a molecule-molecule bound state, but rather a resonance between two free scattering states enabled by the internal structure of the molecules (*39*).

The experimental setup has been described in detail previously (*31*). In brief, a degenerate mixture of $^{40}$K and $^{87}$Rb is prepared in 6 layers of a 1D optical lattice, with final trap frequencies $(\omega_x, \omega_y, \omega_z) = 2\pi \times (34,\ 17.7 \times 10^3,\ 34)$ Hz for KRb in each layer (gravity points along $-\hat{y}$). Weakly-bound molecules are created with a magnetic field ramp through an interspecies Feshbach resonance at 546.62 G, and are transferred to the rovibronic ground state via stimulated Raman adiabatic passage (STIRAP) at $|\vec{E}_{\text{STIRAP}}| = 4.5$ kV/cm with the field along $+\hat{y}$. For studying the $|1,0\rangle$ state, it is advantageous to perform STIRAP at large $|\vec{E}|$ to bypass several avoided crossings at $|\vec{E}| < 1$ kV/cm that arise from the hyperfine structure (*23*). Typical starting conditions are $2 \times 10^4$ molecules in the $|0,0\rangle$ state at a temperature $T = 250$ nK, corresponding to about 1.8 times the Fermi temperature. With nearly perfect occupancy of the lowest band ($\frac{k_B T}{\hbar \omega_y} \sim 0.3$, with $k_B$ the Boltzmann constant and $\hbar$ the reduced Planck constant) and negligible tunneling between lattice sites, our system realizes a stack of quasi-2D molecular gases.

To measure the reactive loss of the $|1,0\rangle$ state, we use the following protocol. Starting at $\vec{E}_{\text{STIRAP}}$, we first apply a microwave $\pi$-pulse to transfer the molecules from $|0,0\rangle$ to $|1,0\rangle$ with a Rabi frequency of $2\pi \times 200$ kHz and a typical efficiency above 95%. Any remaining $|0,0\rangle$ population is quickly lost in a few milliseconds via *s*-wave reactive collisions with $|1,0\rangle$. Next, $\vec{E}$ is ramped to its target configuration in 60 ms. After a variable hold time $t$, $\vec{E}$ is ramped back to $\vec{E}_{\text{STIRAP}}$ in 60 ms. To image $|1,0\rangle$ molecules, we apply another microwave pulse to transfer the molecules back to the $|0,0\rangle$ state, then use STIRAP to transfer to the Feshbach state before imaging the molecules in time-of-flight expansion. We fit the measured average density $n$ as a function of $t$ to the solution of the two-body loss rate equation $\frac{dn}{dt} = -\beta n^2$, where $\beta$ is the two-body chemical reaction rate coefficient. In 2D, there is no temperature increase associated with the two-body loss (*31*, *47*), hence this rate equation is simplified in comparison to 3D (*29*).

To fully characterize the shielding effect, we first show how to tune the angular momentum projection $M$ of the collisions by changing the orientation of $\vec{E}$ relative to the quasi-2D planes. The two-body loss process depends on how the two reactants approach relative to the direction of the induced dipole, or more formally, on the collisional partial wave of the relative motion of the two particles with respect to the quantization axis defined by $\vec{E}$. Previous measurements in quasi-2D with $\vec{E}$ oriented along the tightly confined direction ($\hat{y}$) showed a suppression of $\beta$ at moderately large values of $d$ owing to repulsive $M = \pm 1$ dipolar collisions (*31*, *37*). Here, we study the dipolar anisotropy by tilting $\vec{E}$ away from the *y*-axis by an angle $\theta$, which controllably mixes in the attractive $M = 0$ scattering channel (Fig. 2A). Note that the coupling to $\vec{E}$ is by far the dominant perturbation to the rotational energies, and accordingly, $M$ is always defined relative to the axis of $\vec{E}$. While the dipolar interaction in general mixes higher partial waves into the scattering, we expect that the contributions from $|M| > 1$ are negligible for the relatively small values of $d$ explored here (*48*).

While the collisions always occur along the $\hat{x}$ and $\hat{z}$ directions owing to the strong confinement along $\hat{y}$, the angular momentum character of the collisions with respect to $\vec{E}$ changes with $\theta$. For $\theta = 0°$, the collisions decompose equally into the $M = \pm 1$ channels, which give equal contributions to the collision cross section owing to the azimuthal symmetry of the dipolar interaction. For $\theta = 90°$, collisions along $\hat{x}$ correspond to $M = 0$ scattering, while those along $\hat{z}$ are still an equal superposition of the $M = \pm 1$ channels. Hence, by measuring $\beta$ at $\theta = 0°$ and $90°$, one can extract the loss rate coefficients $\beta_{\pm 1}$ and $\beta_0$ associated with the $M = \pm 1$ and 0 channels, respectively. The full dependence on $\theta$, shown in Fig. 2B, can be calculated by considering the mixing of the $M$ states under rotations (*49*).

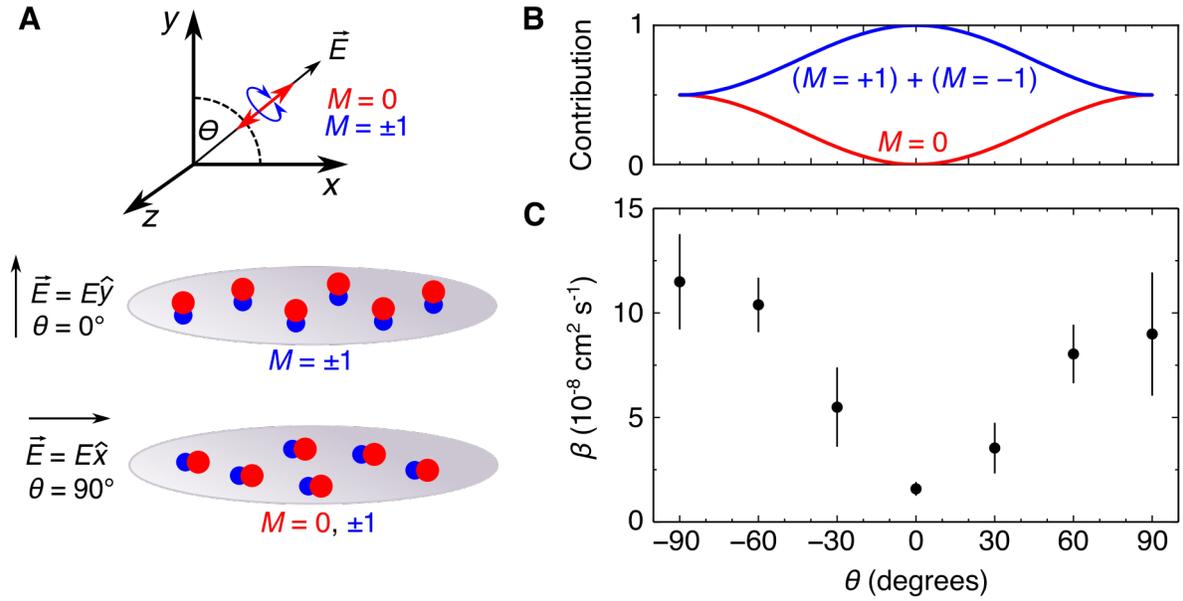

**Fig. 2. Experiment setup and control of reactions by the electric field orientation.**

(**A**) Schematic of the experiment geometry. $\vec{E}$ defines the quantization axis and can be rotated in the $x$-$y$ plane, forming an angle $\theta$ with respect to $y$. A 1D optical lattice confines the molecules to the $x$-$z$ plane. As a result, collisions always occur along $\hat{x}$ and $\hat{z}$, however the angular momentum projection $M$ of the collisions with respect to $\vec{E}$ depends on $\theta$. For $\theta = 0°$, collisions along $\hat{x}$ or $\hat{z}$ occur only in the $M = \pm 1$ projections. For $\theta = 90°$, collisions along $\hat{x}$ ($\hat{z}$) correspond to the $M = 0$ ($\pm 1$) projection. (**B**) Contribution of the $M = 0$ (red) or $\pm 1$ (blue) channels to the scattering as a function of $\theta$. (**C**) $\beta$ as a function of $\theta$ for a fixed bias strength $|\vec{E}| = 7.09$ kV/cm. Error bars are 1 s.e. from fits to the two-body rate equation.

Figure 2C shows the measured $\beta$ for $|1,0\rangle$ molecules at $|\vec{E}| = 7.09$ kV/cm as $\theta$ is varied over 180°. As expected, $\beta$ increases with $|\theta|$ and reaches a maximum at $\theta = \pm 90°$, where attractive $M = 0$ collisions dominate the loss rate. At 7.09 kV/cm, the relatively small value of $d = -0.12$ D limits the maximum increase of $\beta$ to only an order of magnitude, in contrast to the much larger effect expected for larger $d$ (*48*). Our electrode geometry permits excellent control of the curvature of $\vec{E}$ along $\hat{x}$, except near $\theta = \pm 90°$, where we applied a small correction to the measured $\beta$ to account for compression of the cloud owing to the increased curvature in this configuration (*49*).

Having controlled the angular momentum channels participating in the collisions, we proceed to explore the dependence of the $|1,0\rangle$ reaction rate on $\vec{E}$. We measure $\beta$ at both $\theta = 0°$ and 90° to extract $\beta_{\pm 1}$ and $\beta_0$ as a function of $|\vec{E}|$ (*49*), which are summarized in Figure 3A and 3B respectively. For both values of $\theta$, we calibrate $|\vec{E}|$ to a few parts in $10^4$ using spectroscopy on the $|0,0\rangle$ to $|1,0\rangle$ transition.

In the background region ($|\vec{E}| = 1$ to 11 kV/cm) away from resonance, we observe a slight decrease in $\beta_{\pm 1}$ and a corresponding increase in $\beta_0$. $\beta_{\pm 1}$ ($\beta_0$) reaches a minimum (maximum) near $|\vec{E}| = 7$ kV/cm, in agreement with theoretical predictions (solid lines). To understand the trends of $\beta_{\pm 1}$ and $\beta_0$, we note that $|d|$ is non-monotonic in the investigated range of $|\vec{E}|$, and reaches a maximum of 0.12 D at 7 kV/cm. Thus, the trends of $\beta_{\pm 1}$ and $\beta_0$ are consistent with the semiclassical picture of repulsive side-to-side ($M = \pm 1$) or attractive head-to ($M = 0$) dipolar collisions modifying the loss rate, as previously measured for the $N = 0$ state (*31*, *37*, *38*). Away from resonance, our results illustrate the universal nature of this effect for reactive molecules, depending only on the value of $|d|$ and not on the rotational state of the molecule, when resonant dipolar effects are not important.

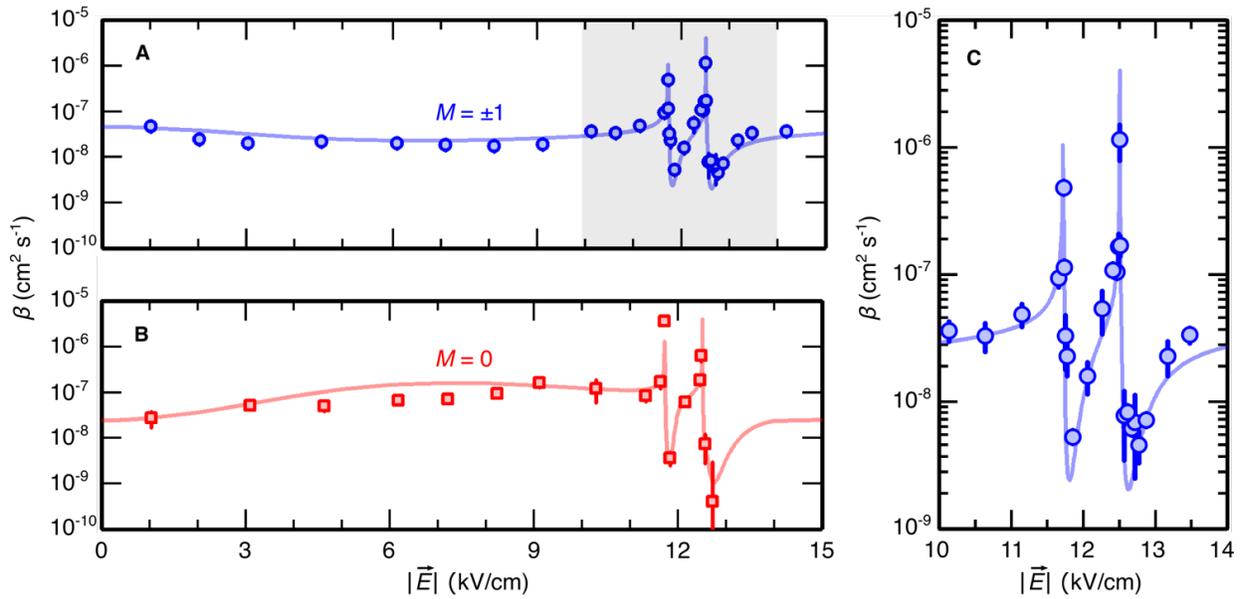

**Fig. 3. 3D characterization of the shielding effect.** (**A**) $\beta_{\pm 1}$ (blue circles) versus $|\vec{E}|$ extracted from loss measurements at $\theta = 0°$. (**B**) $\beta_0$ (red squares) versus $|\vec{E}|$ extracted from loss measurements at $\theta = 90°$ and $0°$. In (A) and (B), solid lines are theoretical predictions for the experimental $T$ and $\omega_y$ with no free parameters. Error bars are 1 s.e. from fits to the two-body loss equation. (**C**) Close-up of $\beta_{\pm 1}$ in the region near $|\vec{E}_1|$ and $|\vec{E}_2|$ (gray shaded region of (A)).

In the region near $|\vec{E}_1|$ and $|\vec{E}_2|$, resonant off-diagonal dipolar couplings to $|0,0\rangle|2,\pm1\rangle$ and $|0,0\rangle|2,0\rangle$ become the dominant contribution, instead of the diagonal dipolar interactions that determine $\beta$ in the background region. We observe sharp features at $|\vec{E}_1|$ and $|\vec{E}_2|$ in both the $M = \pm1$ and $M = 0$ channels, in excellent quantitative agreement with the scattering theory predictions with no free parameters (solid lines) (*49*). We measure a maximum variation of $\beta_{\pm1}$ by a factor of 300(20), and a factor of 8(3) reduction in $\beta_{\pm1}$ at the optimal shielding condition (12.77 kV/cm) compared to the value away from the features (10.13 kV/cm). In the $M = 0$ channel, we observe a variation of $\beta_0$ by a factor of 1000(400) near the features. Comparing the measurements at the optimal shielding point (11.84 kV/cm) and away from the features (11.32 kV/cm), we observe a maximum suppression of $\beta_0$ by a factor of 23(10) below its background value. (We excluded the point at 12.72 kV/cm from this analysis, since the extracted $\beta_0$ is consistent with zero within our measurement precision.)

As opposed to the semi-classical nature of the loss suppression when dipolar molecules are made to collide side-by-side (*31*, *37*), the presence of the resonances in both the $M = 0$ and $\pm1$ channels highlights the quantum nature of the shielding mechanism, which is based on level repulsion between the two scattering channels brought to degeneracy by $\vec{E}$. The effect occurs independently of the relative orientation of the dipoles, since approaching with either $M = 0$ or $\pm1$ causes dipolar mixing of the rotational channels. Our quasi-2D measurements indicate that the shielding is present in all three of these channels at the same values of $|\vec{E}|$, even in the case of $M = 0$ where the dipoles approach in attractive head-to-tail collisions. We thus fully expect that the shielding will be effective also in 3D (*39*).

Figure 3C shows a close-up of $\beta_{\pm1}$ near $|\vec{E}_1|$ and $|\vec{E}_2|$, emphasizing the narrow widths of the features. An intuitive explanation for the observed widths, which are on the order of tens of V/cm, comes from comparing the resonant dipolar interaction energy with the Stark shift of the two crossing channels. Away from resonance, the reaction rate is controlled by the height of the $p$-wave centrifugal barrier, which occurs at $r_0$ (*43*). Near resonance, this barrier is modified by the dipolar coupling $V_{dd}$ between the channels, with an approximate energy scale of $V_{dd} \sim \frac{d_0^2}{4\pi\epsilon_0 r_0^3} = h \times (16 \text{ MHz})$, where $d_0 = 0.574$ D is the permanent dipole moment, $\epsilon_0$ is the permittivity of free space, and $h$ is the Planck constant. The differential Stark shift of the two channels near 12 kV/cm is roughly $\frac{\partial U}{\partial |\vec{E}|} = h \times 215$ kHz/(V/cm). This suggests a width on the order of $V_{dd}/\left(\frac{\partial U}{\partial |\vec{E}|}\right) = 75$ V/cm, in qualitative agreement with the exact result from scattering calculations.

To underline the dramatic change in $\beta$ under a small variation of $|\vec{E}|$, Figure 4 displays two molecular loss curves in the vicinity of $|\vec{E}_2|$ at $\theta = 0°$, corresponding to the largest measured difference in $\beta_{\pm 1}$. The values of $|\vec{E}|$ for the two curves, 12.50 (green squares and inset) and 12.67 (orange circles) kV/cm, correspond to the adiabatic energy curves in Figure 1C. At $|\vec{E}| = 12.50$ kV/cm, the loss rate is strongly enhanced and the molecules are lost within 100 ms. In contrast, at $|\vec{E}| = 12.67$ kV/cm, the molecules are shielded from loss and ~ 20% of the initial density can still be detected after 20 s of hold time, realizing a long-lived gas of polar molecules in a large electric field.

We have demonstrated a method for controlling reactive losses using an external electric field and find excellent agreement with theoretical predictions. From our investigation of the $M = 0$ and $\pm 1$ collision channels, we fully expect that the shielding remains effective in 3D geometry, without the need for an optical lattice to protect the molecules. The shielding could be used to create a favorable ratio of elastic to inelastic collisions for evaporative cooling, simplifying future efforts to create quantum-degenerate molecular gases for species other than KRb (*41*, *42*). These results provide long-lived quantum gases of polar molecules in strong electric fields that are ready to be used to explore a wide range of exciting many-body phenomena and quantum information applications.

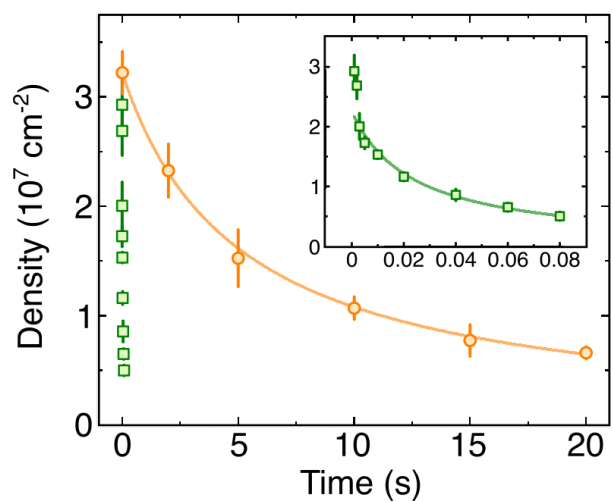

**Fig. 4. Reaction shielding and enhancement near $|\vec{E}_2|$.** Molecular loss measurements at $|\vec{E}| = 12.50$ kV/cm (green squares) and 12.67 kV/cm (orange circles) for $\theta = 0°$. The inset shows the data for $|\vec{E}| = 12.50$ kV/cm on an enlarged x-axis. Solid lines are fits to the two-body rate equation, and error bars are 1 s.e. of independent measurements.

**Acknowledgments:** We thank J. L. Bohn for stimulating discussions and careful reading of the manuscript.

Funding: We acknowledge funding from NIST, DARPA DRINQS, ARO MURI, and NSF Phys-1734006. G.Q. acknowledges funding from the FEW2MANY-SHIELD Project No. ANR-17-CE30-0015 from Agence Nationale de la Recherche.

Author contributions: The experimental work and data analysis were done by K.M., L.D.M., J.-R.L., W.G.T., G.V., and J.Y. Theoretical calculations were done by G.Q. All authors contributed to interpreting the results and writing the manuscript.

Competing interests: The authors declare that they have no competing financial interests.

* Correspondence to: K.M. (kyle.matsuda@colorado.edu) and J.Y. (ye@jila.colorado.edu).

**Supplementary Materials**

Quantum scattering calculation

The theoretical collisional formalism was developed and explained in more detail in Ref. (*39*). To sum up, the calculation was performed using a time-independent quantum formalism in a free space 3D geometry, including the internal rotational structure of the KRb molecule and an external electric field. A usual partial-wave expansion of the colliding wavefunction was employed leading to a set of differential coupled equations. Matching with asymptotic boundary conditions, the scattering matrix was obtained and the 3D rate coefficient for any initial state of the molecules was extracted for a given collision energy and electric field. To account for chemical reaction losses of KRb molecules, a short-range boundary condition is introduced so that when the molecules meet in this region, they are lost with a unit probability, consistent with previous experimental data (*27, 29, 34*). For identical KRb molecules in indistinguishable states, three components of the 3D rate are computed, for $M = 0, \pm 1$. To obtain the corresponding 2D rate coefficient of each component in a quasi-2D geometry, we follow Ref. (*37*) and references therein. We divide the 3D rate coefficient by $\sqrt{\pi} a_{\text{ho}}$, where $a_{\text{ho}} = \sqrt{\hbar/m_{\text{red}}\omega_y}$ is the characteristic length of the harmonic oscillator confinement, with $m_{\text{red}}$ the reduced mass of two colliding KRb molecules. For $\frac{\omega_y}{2\pi} = 17.7$ kHz, we used $a_{\text{ho}} = 94.8$ nm.

Contribution of $M$ channels versus $\theta$

The angular momentum projections of collisions in the $x$-$z$ plane change as the electric field angle $\theta$ is rotated. For strong dipolar interactions, higher partial waves are mixed into the collisions and thus higher $|M|$ states must be included, in which case a rigorous treatment based on tesseral harmonics can be used (*48*). For the relatively small values of $|d| < 0.2$ D studied here, it is a good approximation to consider only the $M = 0, \pm 1$ projections.

To calculate the contributions of the three $M$ states to the scattering, we start with a 3D picture of identical fermionic molecules scattering in the lowest $L = 1$ partial wave, where $L$ is the total angular momentum of the relative motion between the colliding molecules and $m_L = M$ is the projection onto the axis of $\vec{E}$. We associate a reactive rate coefficient $\beta_M$ to each projection $M$, and assume that the total $\beta$ is given by the sum of the $\beta_M$'s weighted by probability of scattering with each $M$. We assume that in quasi-2D, these weights are given by the decomposition of collisions along $\hat{x}$ and $\hat{z}$ into those with definite $M$. The change in this decomposition as a function of $\theta$ can be calculated using Wigner rotation matrices.

Calculating these weights gives the following equation for the experimentally measured $\beta$ as a function of $\theta$ in terms of $\beta_0$, $\beta_{+1}$, and $\beta_{-1}$:

$$\beta = f(\theta)\beta_0 + g(\theta)\beta_{+1} + h(\theta)\beta_{-1}, \tag{S1}$$

with $f(\theta) = \sin^2\theta$ and $g(\theta) = h(\theta) = \frac{1}{2}(1 + \cos^2\theta)$. The quantities plotted in Fig. 2B of the main text are $f(\theta)/2$ (red line) and $(g(\theta) + h(\theta))/2$ (blue line), which are scaled so that they sum to 1. We note that the two $|M| = 1$ channels contribute equally to the scattering ($\beta_{+1} = \beta_{-1}$) by the symmetry of the dipolar interaction, so we also define $\beta_{\pm 1} \equiv \beta_{+1} + \beta_{-1} = 2\beta_{+1}$.

At $\theta = 0°$, eq. (S1) gives $\beta = \beta_{\pm 1}$. At $\theta = 90°$, eq. (S1) gives $\beta = \beta_0 + \frac{1}{2}\beta_{\pm 1}$. By measuring $\beta$ at $\theta = 0°$ and $90°$, one obtains measurements of both $\beta_0$ and $\beta_{\pm 1}$.

Tilted field configurations

The in-vacuum electrode assembly has been discussed in detail previously (*31*). In brief, the assembly consists of two ITO coated transparent glass plates and four tungsten rods. We model the electric potential produced by each electrode in the $x$-$y$ plane using a finite-element simulation. In practice, we use a simplified model of the electric field obtained by fitting the potential generated by each electrode with a polynomial up to $x^7 y^7$ order. We have demonstrated the accuracy of this model by measuring the force of the electric field on the molecules as a function of the electrode voltages (see the Methods section of Ref. (*31*)).

Using this simplified model, we can optimize the electrode values to produce a field with the desired $|\vec{E}|$, $\theta$, $d|\vec{E}|/dx$, $d|\vec{E}|/dy$, $d^2|\vec{E}|/dx^2$, and $d^2|\vec{E}|/dy^2$ at the geometrical center of the electrode assembly. In practice, the six electrodes do not provide enough degrees of freedom to constrain all six of these parameters except in cases of high symmetry (e.g., $\theta = 0°$), so for the tilted field measurements with $\theta \neq 0°$, we ignore $d^2|\vec{E}|/dy^2$ in the optimization. The resulting field curvature along $\hat{y}$ corresponds to a harmonic potential on the order of tens of Hz, which is negligible compared to our lattice frequency of 17.7 kHz. The field curvature along $\hat{x}$ is negligible for all field configurations discussed except for $\theta = 90°$, which is discussed in detail below.

Electric field curvature at $\theta = 90°$

At $\theta = 90°$, the field is produced using only the rods (with the plates grounded), owing to the electrode geometry. This creates a significant field curvature along $\hat{x}$, which changes the combined trapping frequency of the electrical and optical potentials by up to 50% (Fig. S1). While the molecules are always imaged in the same condition (at $\vec{E}_{\text{STIRAP}}$ with $\theta = 0°$, with negligible electric field curvature), the large curvature at $\theta = 90°$ can compress the cloud and raise the density during the hold time, leading to an apparently higher measured $\beta$.

We correct our measured values of $\beta$ to account for this effect by assuming that the cloud is adiabatically compressed as $\vec{E}$ is ramped to its target value. From the measured molecule density $n$, the trap frequency $\omega_{x,0} = 2\pi \times 31.2$ Hz at $\vec{E}_{\text{STIRAP}}$, and the calculated trap frequency $\omega_x(|\vec{E}|)$ at $\theta = 90°$ as a function of $|\vec{E}|$, we can estimate the density $n(|\vec{E}|)$ during the hold time,

$$n(|\vec{E}|) = n \times \left(\frac{\omega_x(|\vec{E}|)}{\omega_{x,0}}\right)^{1/2}. \qquad (S2)$$

The measurements of $\beta_0$ reported in Fig. 3B of the main text are obtained using the corrected densities from eq. (S2). The correction on $\beta$ depends on $|\vec{E}|$, but is at most 30% of the measured value.

In principle, the cloud is also adiabatically heated by the additional electric field curvature. This may cause an additional systematic effect arising from the dependence of $\beta$ on $T$. We do not correct for this effect, but we note that the correction be of the same order as the correction on $n$, assuming the Wigner threshold law scaling $\beta \sim T$. Owing to the slow variation in $\omega_x(|\vec{E}|)$ near the resonances, we expect that neither the adiabatic compression nor heating has a significant effect on the observed variation of $\beta$ near the resonances.

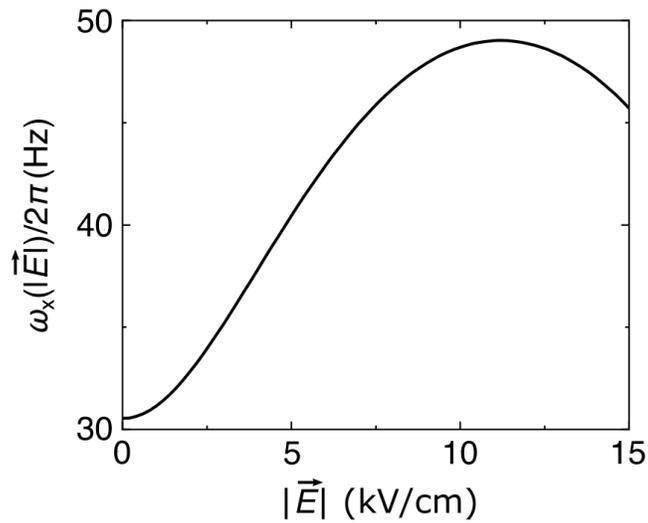

**Fig. S1. Electric field curvature at $\theta = 90°$.** The calculated trap frequency $\omega_x(|\vec{E}|)$ of $|1,0\rangle$ molecules in the combined electric and optical potentials.